\documentclass[a4paper,11pt]{article}
\pdfoutput=1 
\usepackage{jcappub}
\usepackage{amsmath}
\usepackage{bm}
\usepackage{graphicx}
\usepackage{enumitem}
\usepackage{geometry}
\usepackage{xspace}
\usepackage{pdflscape}

\makeatletter
\gdef\@fpheader{}
\g@addto@macro\bfseries{\boldmath}
\makeatother

\newcommand{\ie}{{i.e.~}}

\newcommand{\eg}{e.g.~}

\let\oldsqrt\sqrt
\def\sqrt{\mathpalette\DHLhksqrt}
\def\DHLhksqrt#1#2{%
\setbox0=\hbox{$#1\oldsqrt{#2\,}$}\dimen0=\ht0
\advance\dimen0-0.2\ht0
\setbox2=\hbox{\vrule height\ht0 depth -\dimen0}%
{\box0\lower0.4pt\box2}}

\newcommand{\dd}{\mathrm{d}}
\newcommand{\ee}{e}

\newcommand{\sss}[1]{{\scriptscriptstyle{#1}}}

\newcommand{\SR}{{{}_\mathrm{SR}}}
\newcommand{\USR}{{{}_\mathrm{USR}}}

\newcommand{\uPl}{\mathrm{Pl}}
\newcommand{\uin}{\mathrm{in}}

\newcommand{\umax}{\mathrm{max}}

\newcommand{\usssPl}{\sss{\uPl}}

\newcommand{\Mp}{M_\usssPl}

\newcommand{\efolds}{$e$-folds}

\newcommand{\beq}{\begin{equation}}
\newcommand{\eeq}{\end{equation}}
\newcommand{\bea}{\begin{equation}\begin{aligned}}
\newcommand{\eea}{\end{aligned}\end{equation}}

\newlength{\wsingfig}
\newlength{\wdblefig}
\newlength{\wquadfig}
\newlength{\wtriplefig}
\newcommand{\Eq}[1]{Eq.~(\ref{#1})}
\newcommand{\Eqs}[1]{Eqs.~(\ref{#1})}
\newcommand{\Fig}[1]{Fig.~{\ref{#1}}}
\newcommand{\Figs}[1]{Figs.~{\ref{#1}}}
\newcommand{\Ref}[1]{Ref.~{\cite{#1}}}

\newcommand{\Sec}[1]{Sec.~\ref{#1}}

\subheader{}

\title{The attractive behaviour of ultra-slow-roll inflation}

\author[a]{Chris Pattison,}
\author[b,a]{Vincent Vennin,}
\author[a,c]{Hooshyar Assadullahi,}
\author[a]{and David Wands}

\affiliation[a]{Institute of Cosmology \& Gravitation, University of Portsmouth, Dennis Sciama Building, Burnaby Road, Portsmouth, PO1 3FX, United Kingdom}
\affiliation[b]{Laboratoire Astroparticule et Cosmologie, Universit\'e Denis Diderot Paris 7, 75013 Paris,
France}
\affiliation[c]{School of Mathematics and Physics, University of Portsmouth, Lion Gate Building, Lion Terrace, Portsmouth, PO1 3HF, United Kingdom}

\emailAdd{christopher.pattison@port.ac.uk}
\emailAdd{vincent.vennin@port.ac.uk}
\emailAdd{hooshyar.assadullahi@port.ac.uk}
\emailAdd{david.wands@port.ac.uk}

\date{today}

\begin{document}

\sloppy

\abstract{It is often claimed that the ultra-slow-roll regime of inflation, where the dynamics of the inflaton field are friction dominated, is a non-attractor and/or transient. In this work we carry out a phase-space analysis of ultra-slow roll in an arbitrary potential, $V(\phi)$. We show that while standard slow roll is always a dynamical attractor whenever it is a self-consistent approximation, ultra-slow roll is stable for an inflaton field rolling down a convex potential with $\Mp V''>|V'|$ (or for a field rolling up a concave potential with $\Mp V''<-|V'|$). In particular, when approaching a flat inflection point, ultra-slow roll is always stable and a large number of \efolds~may be realised in this regime. However, in ultra-slow roll, $\dot{\phi}$ is not a unique function of $\phi$ as it is in slow roll and dependence on initial conditions is retained. We confirm our analytical results with numerical examples.}



\maketitle

\section{Introduction}
\label{sec:intro}

Cosmological inflation \cite{Starobinsky:1980te, Sato:1980yn, Guth:1980zm, Linde:1981mu, Albrecht:1982wi, Linde:1983gd} is the leading paradigm of very early Universe physics, and describes a period of accelerated expansion immediately after the Big Bang. 
It successfully solves many problems of standard hot big bang cosmology, and also provides seeds (in the form of quantum fluctuations) that are amplified to become the large-scale structure we see today \cite{Mukhanov:1981xt, Mukhanov:1982nu, Starobinsky:1982ee, Guth:1982ec, Hawking:1982cz, Bardeen:1983qw}.

The simplest realisation of inflation is when the accelerated expansion is driven by a single scalar field $\phi$, called the inflaton. Its classical equation of motion in a Friedmann-Lema\^itre-Robertson-Walker (FLRW) cosmology is given by the Klein-Gordon equation
\bea
\ddot{\phi} + 3 H\dot{\phi} + V'(\phi) = 0\, ,
\label{eq:eom:kleingordon}
\eea
where a dot denotes a derivative with respect to cosmic time $t$, $V'$ is the derivative of the potential with respect to the inflaton field value, and $H=\dot{a}/a$ is the Hubble expansion rate ($a$ being the scale factor), that satisfies the Friedmann equation
\bea
\label{eq:friedmann}
H^2 = \frac{1}{3\Mp^2} \left( V+\frac{\dot{\phi}^2}{2} \right) \, ,
\eea
where $\Mp=(8\pi G_{\mathrm{N}})^{-1/2}$ is the reduced Planck mass. It is generically not possible to solve this system of equations analytically, and so approximations are often made to simplify the dynamics. 

The most common approximation is that of slow-roll (SR) inflation.
In this regime, one takes $\dot{\phi}^2 \ll V(\phi)$, so the energy budget of the inflaton is potential dominated and \Eq{eq:friedmann} reads $H^2\simeq V/(3\Mp^2)$. Under this approximation, one also neglect the acceleration term $\ddot{\phi}$ in the Klein-Gordon equation~(\ref{eq:eom:kleingordon}), and hence the equation of motion becomes first order,
\bea 
\label{eq:eom:KG:sr}
\dot{\phi}_\SR\simeq -\frac{V'}{3H} \, .
\eea

However, there are regimes in which this approximation is not a valid one.
For instance, one can imagine cases where the potential of the inflaton becomes very flat, so $V'(\phi) \to 0$, and SR begins to break down. 
A simple example of this is a potential with a flat infection point at which $V' =V''= 0$.
Under the SR approximation, \Eq{eq:eom:KG:sr} would give us $\dot{\phi} = 0$ at the inflection point, and so the inflaton rolling down this potential would come to a complete stop at the flat point of the potential.
In practice, we may expect the residual (although small) kinetic energy to carry the field through the inflection point, contrary to the SR prediction.
This motivates us to study these very flat regions in the potential, when SR is violated and a phase of so-called ``ultra-slow-roll" (USR)~\cite{Inoue:2001zt, Kinney:2005vj}, or ``friction dominated", inflation takes place. 

Let us take a moment at this point to clear up the nomenclature surrounding ``ultra-slow-roll" inflation and similar scenarios, as this can sometimes be unclear in the literature. 
``Ultra-slow roll" is the name we use to describe the situation when the potential of the inflaton is very flat, so that $V' \simeq 0$, and hence from the Klein-Gordon equation~(\ref{eq:eom:kleingordon}) 
\bea
\label{eq:eom:KG:usr}
\ddot{\phi}_\USR\simeq-3H\dot{\phi}_\USR\, .
\eea
``Constant-roll" inflation~\cite{Martin:2012pe, Motohashi:2014ppa, Karam:2017rpw, Yi:2017mxs, Morse:2018kda} is intended to be a generalisation of ultra-slow-roll inflation, and is simply defined as a regime where $-\ddot{\phi}/(3H\dot{\phi}) = \mathrm{constant}$, and this constant is not necessarily equal to $1$.
However, often when constant-roll inflation is considered, ultra-slow roll is actually a singular point of the equations studied, and so the analysis does not include ultra-slow roll. One must thus be careful when calling constant-roll inflation a generalisation of ultra-slow roll. Another situation where SR is violated is ``fast-roll" inflation \cite{Linde:2001ae}, where the effective mass of the inflaton is of the same order as $H$ and all three terms in the Klein-Gordon equation~(\ref{eq:eom:kleingordon}) are of comparable magnitude (in the ``ultra-fast roll'' limit, the friction term is subdominant).

In this work, we are interested in the stability properties of USR inflation. While SR inflation is known to be a dynamical attractor~\cite{Salopek:1990jq, Liddle:1994dx, Vennin:2014xta, Grain:2017dqa}, it is often claimed in the literature that USR is always non-attractive~\cite{Cai:2017bxr, Dimopoulos:2017ged, Anguelova:2017djf, Biagetti:2018pjj, Morse:2018kda}. However, this conclusion is obtained from investigating constant-roll inflation, which is supported only by a very specific class of potentials and which, as already pointed out, only reduces to USR in a (singular) limit. This is why we carry out a generic analysis of USR that does not assume a specific potential. We will derive a simple criterion on the potential for when it is stable.

This paper is arranged as follows. In \Sec{sec:d&d} we introduce our notations and discuss the SR and USR regimes of the inflaton dynamics. In \Sec{sec:stability} we discuss the possibility of a stable USR period and derive a necessary and sufficient condition for stability in USR. We then demonstrate how this condition can be applied to some simple examples in \Sec{sec:examples}, and conclude in \Sec{sec:conclusions}.
\section{Inflaton dynamics and definitions}
\label{sec:d&d}
We begin by discussing the formalism and language we will use to construct and define USR, and explain how it differs from SR.
\subsection{Hubble-flow parameters}
Before making any approximations, let us first introduce the set of Hubble-flow parameters. Starting from $\epsilon_0 \equiv H_\uin/H$, one can define 
\bea
\label{def:epsilonn}
\epsilon_{n+1} \equiv \frac{\dd\ln(\epsilon_n)}{\dd N} \, ,
\eea
where $N\equiv \ln a = \int H\, \dd t$ is the number of \efolds . 
Inserting the Klein-Gordon equation~(\ref{eq:eom:kleingordon}) in the time derivative of the Friedmann equation~(\ref{eq:friedmann}), one obtains the first dimensionless Hubble-flow parameter
\beq
\label{eq:eps1exact}
\epsilon_1\equiv -\frac{\dot{H}}{H^2}=3\frac{\dot{\phi}^2/2}{V+\dot{\phi}^2/2}\, .
\eeq
The condition for inflation, $\ddot{a}>0$, corresponds to $\epsilon_1<1$ since $\epsilon_1=1-a \ddot{a}/\dot{a}^2$. Inserting the Klein-Gordon equation~(\ref{eq:eom:kleingordon}) in the time derivative of \Eq{eq:eps1exact}, one obtains the second dimensionless Hubble-flow parameter
\bea
\label{eq:eps2exact}
\epsilon_2\equiv \frac{\dot{\epsilon_1}}{H\epsilon_1} =
6\left(\frac{\epsilon_1}{3}-\frac{V^\prime}{3H\dot{\phi}}-1\right)\, .
\eea
In general, there is no requirement that $\epsilon_2$ is small during inflation, in contrast to $\epsilon_1$.
\subsection{Field acceleration parameter}
We now introduce the field acceleration parameter $f$, that quantifies the relative importance of the acceleration term compared with the friction term in the Klein-Gordon equation~(\ref{eq:eom:kleingordon}),
\bea
\label{eq:def:f}
f\equiv -\frac{\ddot{\phi}}{3H\dot{\phi}}
=1+\frac{V'}{3H\dot{\phi}}
\ .
\eea
In the second equality of the above expression, we have used the Klein-Gordon equation~(\ref{eq:eom:kleingordon}). The field acceleration parameter can be expressed in terms of the first two Hubble-flow parameters, as can be seen from combining \Eqs{eq:eps1exact} and \eqref{eq:eps2exact} with \Eq{eq:def:f},
\bea
\label{eq:f:epsilon}
f = \frac{2\epsilon_1-\epsilon_2}{6}\, .
\eea

Since $f$ is a function of $\phi$ and $\dot{\phi}$, phase space (which is usually parametrised by $\phi $ and $\dot{\phi}$) can also be parametrised by $\phi$ and $f$. This will prove useful in the following. To this end, let us express $\dot{\phi}$ in terms of $\phi$ and $f$, which can be done by combining \Eqs{eq:friedmann} and~(\ref{eq:f:epsilon}),
\bea
\label{eq:phidot:f:phi}
\dot{\phi}^2 = V\left[\sqrt{1+\frac{2\Mp^2}{3\left(f-1\right)^2}\left(\frac{V'}{V}\right)^2}-1\right]\, .
\eea
Thus by combining \Eqs{eq:eps1exact} and~(\ref{eq:phidot:f:phi}), one can write the first Hubble flow parameter as
\bea
\label{eq:eps1:f:phi}
\epsilon_1 =  3\frac{\sqrt{1+\frac{2\Mp^2}{3\left(1-f\right)^2}\left(\frac{V'}{V}\right)^2}-1}{\sqrt{1+\frac{2\Mp^2}{3\left(1-f\right)^2}\left(\frac{V'}{V}\right)^2}+1}\, .
\eea
The condition for inflation to take place, $\epsilon_1<1$, then reads
\bea
\label{eq:inflation:condition}
\frac{\Mp}{\left\vert 1-f \right\vert }\left\vert \frac{V'}{V}\right\vert < \frac{3}{\sqrt{2}}\, .
\eea
We will often work in the quasi-de Sitter (quasi-constant-Hubble) approximation which corresponds to $\epsilon_1\ll1$, and hence $[\Mp/\vert 1-f\vert]\vert {V'}/{V} \vert\ll1$.
\subsection{Slow-roll inflation}
Slow-roll inflation corresponds to the regime where all Hubble flow parameters are much smaller than one, \ie $\vert \epsilon_n\vert \ll 1$ for $n\geq 1$. From \Eq{eq:eps1exact}, this means that the kinetic energy of the inflaton field is much smaller than its potential energy, and from \Eq{eq:def:f}, it implies that the acceleration of the inflaton field can be neglected\footnote{In this sense SR corresponds to quasi-equilibrium, \ie zero net force with frictional force equal and opposite to the force from the potential gradient.} compared with its friction, so the dynamical system boils down to
\bea
\label{eq:slowroll}
H_\SR^2 \simeq \frac{V}{3\Mp^2}\,  \quad {\rm and} \quad 3H\dot\phi_\SR \simeq - V' \, ,
\eea
and hence $|f|\ll 1$.
In this limit, $\dot{\phi}$ is determined completely by the gradient of the potential and a single trajectory is selected out in phase space since $\dot{\phi}_\SR$ has no dependence on initial conditions. One notices that while SR is usually defined as $\vert\epsilon_n\vert \ll 1$ for all $n\geq 1$, the above system only relies on $\epsilon_1\ll 1$ and $\vert \epsilon_2 \vert \ll 1$.

Since $\dot{\phi}$ is an explicit function of $\phi$ through \Eq{eq:slowroll}, any phase space function can be written as a function of $\phi$ only. For the first Hubble-flow parameter and the field acceleration parameter, substituting \Eq{eq:slowroll} into \Eqs{eq:eps1exact} and \eqref{eq:def:f}, one obtains
\bea
\epsilon_{1\SR} &\simeq \frac{\Mp^2}{2}\left(\frac{V'}{V}\right)^2 \,,
\eea
\bea
f_\SR &\simeq \frac{\Mp^2}{3} \left[\frac{V''}{V} - \frac{1}{2}\left(\frac{V'}{V}\right)^2\right]\, .
\label{eq:f:SR}
\eea
For a given inflationary potential $V(\phi)$, the existence of a regime of SR inflation can thus be checked by verifying that  the potential slow-roll parameters $\epsilon_V$ and $\eta_V$, defined as
\bea
\label{eq:epsV}
\epsilon_{V} \equiv \frac{\Mp^2}{2}\left(\frac{V'}{V}\right)^2 
\quad {\rm and} \quad
\eta_V \equiv \Mp^2 \frac{V''}{V} \, ,
\eea
remain small,
\bea
\label{eq:sr:consistency}
 \epsilon_{V} \ll 1 \quad {\rm and} \quad |\eta_V |\ll1 \,.
\eea
\subsection{Ultra-slow roll inflation}
\label{sec:Def:USR}

In the ultra-slow roll regime it is the driving term, corresponding to the gradient of the potential, that is neglected in the Klein-Gordon equation \eqref{eq:eom:kleingordon}, rather than the field acceleration. This corresponds to the relative field acceleration $f\approx 1$ in \Eq{eq:def:f}, leading to \Eq{eq:eom:KG:usr}. Note that if $\phi$ follows the gradient of its potential, then $\dot{\phi} V' = \dot{V}<0$ and conversely $\dot{\phi} V'>0$ if the field evolves in the opposite direction. From \Eq{eq:def:f}, one can thus see that $f<1$ corresponds to situations where the inflaton rolls down its potential and $f>1$ to cases where the field climbs up its potential.

Integrating \Eq{eq:eom:KG:usr} leads to the USR solution
\bea
\label{eq:USR:phidot:N}
\dot{\phi}_{\USR} \propto \ee^{-3N} \, .
\eea
This is the USR limit in which one takes $V' = 0$, but we shall see later that other solutions exist approaching USR when $V'$ does not exactly vanish.
Instead of being driven by $V'$ as in the SR case \eqref{eq:eom:KG:sr}, here the time derivative $\dot{\phi}_{\textrm{usr}}$ is exponentially decreasing with the number of \efolds. If we also assume quasi-de Sitter ($\epsilon_1\ll1$), the above can be integrated as
\bea
\label{eq:USR:traj}
\phi_\USR - \phi_{\USR,*} \simeq \frac13 \frac{\dot{\phi}_{\USR,*}}{H_*}  \left[ 1-\ee^{-3\left(N-N_*\right)} \right]  \, ,
\eea
where the star denotes some reference time. Thus the USR solution may be thought of as the free or transient response of the scalar field in an expanding FRLW cosmology. It is independent of the shape of the potential, but depends instead on the initial value of the field and its time derivative, $\phi_*$ and $\dot\phi_*$.

Despite the different background evolution, linear fluctuations of a massless field about ultra-slow-roll inflation have the same scale-invariant form as during slow-roll inflation \cite{Seto:1999jc,Leach:2001zf,Kinney:2005vj}. This is a striking example of the invariance of field perturbations under ``duality'' transformations \cite{Wands:1998yp,Biagetti:2018pjj}. 

The condition under which USR takes place reads $\vert f-1 \vert \ll 1$, which implies that $3H \vert\dot{\phi}\vert \gg \vert V' \vert$. Clearly this is possible for any finite potential gradient so long as we have a sufficiently large field kinetic energy.
However, in order to have inflation we also need to have $\epsilon_{1} < 1$, which from \Eq{eq:inflation:condition} corresponds to
\bea
\label{eq:consistency:usr}
\epsilon_{V} < \frac94 \left(1-f \right)^2
 \, ,
\eea
where $\epsilon_{V}$ is the first potential slow-roll parameter given in \Eq{eq:epsV}. The quasi-de Sitter approximation $\epsilon_{1} \ll 1$ simply corresponds to $\epsilon_{V} \ll (1-f)^2$. Comparing this relation with \Eq{eq:sr:consistency}, one can see that USR inflation requires a potential that is even flatter than what SR imposes at the level of $\epsilon_V$ (hence the name ``ultra''-slow roll, which is otherwise not so apt since SR and USR are disjoint regimes), but that no constraint is required on $\eta_V$, \ie on the second derivative of the potential. 

In the following, we thus distinguish two regimes: USR, that corresponds to $\vert f-1\vert \ll 1$, and USR \emph{inflation}, that corresponds to $\sqrt{\epsilon_V}\ll\vert f-1\vert \ll 1$.
\section{Stability analysis}
\label{sec:stability}
USR is often referred to as a transient or non-attractor solution during inflation~\cite{Cai:2017bxr, Dimopoulos:2017ged, Anguelova:2017djf, Biagetti:2018pjj, Morse:2018kda}. This is because of results in constant-roll models~\cite{Martin:2012pe, Motohashi:2014ppa, Karam:2017rpw, Yi:2017mxs, Morse:2018kda}, where the field acceleration parameter $f$ defined in \Eq{eq:def:f} is taken to be a constant. In the Hamilton-Jacobi formalism, this corresponds to taking $H(\phi)\propto \exp (\pm\sqrt{3f/2}\phi/\Mp)$, and the potentials that support such a phase of constant roll can be obtained from $V=3\Mp^2H^2-2\Mp^4 H'^2$. In these potentials, the constant-roll solution is only one possible trajectory in phase space and one can study its stability. One finds that the constant-roll solution is an attractor if $f<1/2$~\cite{Motohashi:2014ppa}. This excludes the USR limit $f\simeq 1$, which could lead to the incorrect conclusion that USR is always unstable. However this result only applies to the family of potentials mentioned above. Moreover, it is singular in the limit $f \rightarrow 1$ since combining the equations above, one finds $V\equiv \mathrm{constant}$ in that case, for which $f=1$ is the only solution so nothing can be concluded about its attractive or non-attractive behaviour.

This motivates us to go beyond these considerations and to study the phase-space stability of USR in a generic potential.
\subsection{Dynamical equation for the relative field acceleration}
Since the field acceleration parameter, $f$, quantifies the importance of the acceleration term in the Klein-Gordon equation \eqref{eq:eom:kleingordon}, it essentially parameterises whether we are in SR ($\vert f \vert \ll 1$) or USR ($\vert f-1 \vert \ll 1$).
As such, knowing the evolution of $f$ tells us which regime we are in and when we transition from one to the other, and will allow us to study the stability of the two regimes.
As such, we seek a dynamical equation for $f$.

We begin by recasting \Eq{eq:eom:kleingordon} with $\phi$ as the ``time" variable, which reduces the equation to a first-order differential equation, namely
\bea 
\label{eq:kgphi}
\frac{\dd\dot{\phi}}{\dd\phi} +3H + \frac{V'}{\dot{\phi}} = 0 \, ,
\eea 
which can also be written as
\bea
\label{eq:phiddot:f:phi}
\frac{\dd}{\dd\phi}  \left(\dot{\phi}^2\right) = -2 V'\frac{f}{f-1}\, .
\eea
Combined with \Eq{eq:phidot:f:phi} this leads to an equation for the evolution of $f$,
 \bea
 \frac{\dd f}{\dd \phi} = \frac{3}{2\Mp^2}\frac{V}{V'}\left(f-1\right)^2\left(f+1\right)\left[\sqrt{1+\frac{2\Mp^2}{3\left(f-1\right)^2}\left(\frac{V'}{V}\right)^2}-\frac{1-f}{1+f}\right]-\left(1-f\right)\frac{V''}{V'}\, .
 \label{eq:f:dynamical}
 \eea
This can be written in terms of the potential slow-roll parameters \eqref{eq:epsV} as
\bea
\frac{\dd f}{\dd \phi} = \frac{3}{2\Mp}\frac{\left(f-1\right)^2\left(f+1\right)}{\sqrt{2\epsilon_{V}}}\left[\sqrt{1+\frac{4\epsilon_{V}}{3\left(f-1\right)^2}}-\frac{1-f}{1+f}\right]-\frac{\left(1-f\right)\eta_{V}}{2\Mp\sqrt{2\epsilon_{V}}}\, .
\label{eq:f:dynamical:srparams}
\eea
Note that this equation is exact and does not make any assumption about the smallness or otherwise of the slow-roll parameters.
\subsection{Slow-roll limit}
We shall begin our stability analysis by considering the slow-roll case.
If we expand the right-hand side of \Eq{eq:f:dynamical:srparams} to first order in the potential slow-roll parameters \eqref{eq:epsV} and take $f$ to be of first order in the slow-roll parameters as suggested by \Eq{eq:f:SR}, one obtains
\bea
\frac{V'}{V}\frac{\dd f}{\dd \phi} \simeq \frac{1}{2}\left(\frac{V'}{V}\right)^2 - \frac{V''}{V}+\frac{3}{\Mp^2}f\, .
\label{eq:f:dynamical:SR}
\eea
We see that the right-hand side of \Eq{eq:f:dynamical:SR} vanishes for the slow-roll solution \eqref{eq:f:SR}, which is consistent with the fact that $f$ is first order in the slow-roll parameters and $  V'/V \dd f/\dd \phi  \simeq \dd f/\dd N$ is second order in slow roll. 

The stability of the slow-roll solution can then be studied by considering a deviation from \Eq{eq:f:SR} parametrised by
\bea
f \simeq f_{\mathrm{SR}}+\Delta\, .
\label{eq:f:fSR:epsilon}
\eea
In this expression, $f_{\mathrm{SR}}$ is given by \Eq{eq:f:SR} plus corrections that are second order in slow roll and $\Delta$ describes deviations from slow roll that are nonetheless first order in slow-roll parameters or higher. For instance, we imagine that initially, one displaces $f$ from the standard slow-roll expression given in \Eq{eq:f:SR} (\eg by adding another linear combination of some slow-roll parameters) and study how this displacement evolves in time. 
By substituting \Eq{eq:f:fSR:epsilon} into \Eq{eq:f:dynamical:SR}, one obtains
\beq
\frac{V'}{V}\frac{\dd \Delta}{\dd \phi} \simeq \frac{3}{\Mp^2}\Delta\, ,
\eeq
which at leading order in slow roll, using \Eq{eq:slowroll}, can easily be solved to give
\bea
\Delta \simeq \Delta_\uin \exp\left[-3\left(N-N_\uin\right)\right]\, ,
\eea
which is always decreasing as inflation continues. This shows that SR is a stable attractor solution whenever the consistency conditions \eqref{eq:sr:consistency} are satisfied. This is of course a well-known result~\cite{Salopek:1990jq, Liddle:1994dx} but it is interesting to see how it can be formally proven in the formalism employed in this work.
\subsection{Ultra-slow-roll limit}
\label{sec:stability:USR}
In the ultra-slow-roll limit we have $f=1$, which we can readily see is a fixed point of \Eq{eq:f:dynamical} for any potential. We can therefore carry out a generic stability analysis of this fixed point that is valid for any potential. The results will be illustrated with two specific models in \Sec{sec:examples}.

The strategy is to linearise \Eq{eq:f:dynamical} around $f=1$ by parameterising 
\bea
f=1-\delta\, ,
\eea
where we assume $\vert \delta \vert \ll 1$ in order to study small deviations from USR. The only ambiguity is in the argument of the square root in \Eq{eq:f:dynamical}, that reads $1+\epsilon_V/(6\delta^2)$, since both $\epsilon_V$ and $\delta$ are small numbers. However, from \Eq{eq:consistency:usr} and the discussion below it, one recalls that inflation requires $\epsilon_V<9\delta^2/4$, and $\epsilon_V\ll \delta^2$ ensures quasi de-Sitter inflation $\epsilon_1\ll 1$. This is why \Eq{eq:f:dynamical} should be expanded in the USR \emph{inflation} limit $\sqrt{\epsilon_V}\ll\vert\delta\vert\ll 1$,\footnote{\label{footnote:USR:Non:Inflation}An expansion in the USR \emph{non-inflating} limit, $\vert \delta \vert \ll \sqrt{\epsilon_V}$ and $\vert \delta \vert \ll 1$, can also be performed along similar lines.  At linear order in $\delta$, \Eq{eq:f:dynamical} gives rise to
\bea
\frac{\dd \delta}{\dd \phi} \simeq \left[-\frac{\sqrt{6}}{\Mp}\mathrm{sign}\left(V'\delta \right)+\frac{V''}{V'}\right]\delta\, .
\label{eq:f:dynamical:USR:noninflating}
\eea
If the field follows the gradient of its potential, one obtains the stability condition
\bea
\frac{V''}{\left\vert V' \right\vert }>\frac{\sqrt{6}}{\Mp}\, ,
\label{eq:USR:noninflating:stability:condition}
\eea
and conversely, if the field climbs up the potential, one gets $V''/|V'|<-\sqrt{6}/\Mp$. The solution to \Eq{eq:f:dynamical:USR:noninflating} reads
\bea
\delta \simeq \delta_\uin \frac{V'(\phi)}{V'\left(\phi_\uin\right)}\exp\left(\sqrt{6}\frac{|\phi-\phi_\uin|}{\Mp}\right) \, .
\label{eq:delta:sol:USRnoninflating}
\eea
To determine how $\epsilon_1$ varies, one can plug \Eq{eq:delta:sol:USRnoninflating} into \Eq{eq:eps1:f:phi}. One finds that if the field follows the gradient of its potential, then $\epsilon_1$ decreases if $\epsilon_V<3$ and increases otherwise, and it always decreases if the field climbs up its potential. When $\epsilon_1$ decreases, it may become smaller than one at some point, and a phase of USR \emph{inflation} starts, whose stability properties are discussed in the main text.} which gives rise to
\bea
\label{eq:f:dynamical:USR:new}
\frac{\dd\delta}{\dd\phi}\simeq -\frac{3}{\Mp^2}\frac{V}{V'}\delta^2+\frac{V''}{V'}\delta\, .
\eea
The right-hand side of this expression is proportional to $\delta-\eta_V/3$, so which term dominates depends on the magnitude $\vert\delta\vert$ with respect to $\vert\eta_V\vert$. Since $\delta$ must be larger than $\sqrt{\epsilon_V}$, two possibilities have to be distinguished.

\subsubsection{Case $\eta_V^2<\epsilon_V$}
\label{sec:stability:USR:inflation:case1}

In this case the condition for USR inflation, $\sqrt{\epsilon_V}\ll\vert\delta\vert$, guarantees that $\vert\delta\vert \gg \eta_V$ and the first term on the right-hand side of \Eq{eq:f:dynamical:USR:new} dominates,
\bea
\label{eq:f:dynamical:USR:case1}
\frac{\dd\delta}{\dd\phi}\simeq -\frac{3}{\Mp^2}\frac{V}{V'}\delta^2\, .
\eea
As explained at the beginning of \Sec{sec:Def:USR}, if the field follows the gradient of its potential, $f<1$ and $\delta>0$. If $V'>0$ and $\phi$ decreases with time, then from \Eq{eq:f:dynamical:USR:case1} $\delta$ increases with time and USR is unstable. If $V'<0$ and $\phi$ increases with time, then again $\delta$ increases with time and USR is still unstable. Conversely, if we have $f>1$ and $\delta<0$, so that the field climbs up the potential, if $V'>0$ then $\phi$ increases with time and so does $\delta$, so USR is unstable, and if $V'<0$ then $\phi$ decreases with time and USR is still unstable.

We conclude that USR inflation is always unstable in that case.  The example discussed in \Sec{sec:Staro} corresponds to this situation.

\subsubsection{Case $\epsilon_V<\eta_V^2$}
In this case, which term dominates in \Eq{eq:f:dynamical:USR:new} depends on the magnitude of $\delta$. However, strictly speaking, a stability analysis of the fixed point $\delta\sim 0$ should only deal with its immediate neighbourhood, \ie with the smallest possible values of $\vert \delta \vert$ which in this case are smaller than $\vert \eta_V \vert$ (notice that if $\vert \eta_V \vert \gtrsim 1$  this becomes true for all  $\sqrt{\epsilon_V}\ll \vert \delta \vert \ll 1$). The second term in \Eq{eq:f:dynamical:USR:new} then dominates and one has
\bea
\label{eq:f:dynamical:USR:case2}
\frac{\dd\delta}{\dd\phi}\simeq \frac{V''}{V'} \delta\, .
\eea
A similar discussion as in the previous case can be carried out, by first considering the situation where the field follows the gradient of its potential, so $f<1$ and $\delta>0$. If $V'>0$ and $\phi$ decreases with time then $\vert \delta\vert$ decreases if $V''>0$. If $V'<0$ and $\phi$ increases with time then $\vert \delta\vert$ decreases under the same condition 
\bea
V''>0\, .
\eea
Thus we conclude that USR is stable for a scalar field rolling down a convex potential.
Conversely, we find that USR is stable for a scalar field rolling up a concave potential,
$V''<0$. 

The fact that $\vert \delta\vert$ decreases with time is a necessary condition for USR inflation stability but not a sufficient one, since one also has to check that $\vert\delta\vert$ remains much larger than $\sqrt{\epsilon_V}$, \ie that the system remains inflating. To this end, let us notice that \Eq{eq:f:dynamical:USR:case2} can be integrated and gives
\bea
\delta \simeq \delta_\uin \frac{V'(\phi)}{V'\left(\phi_\uin\right)} \, .
\label{eq:delta:sol}
\eea
This confirms that $\vert \delta \vert$ decreases with time when $\vert V' \vert$ decreases. 
Note that, given $\delta=-V'/3H\dot\phi$, this solution corresponds to $H\dot{\phi} =$constant, which differs from the ultra-slow-roll limit \Eq{eq:USR:phidot:N}.
Substituting \Eq{eq:delta:sol} into \Eq{eq:eps1:f:phi} (expanded in the $\sqrt{\epsilon_V}\ll\vert\delta\vert\ll 1$ limit), one obtains
\bea
\label{eq:USR:stable:eps1:appr}
\epsilon_1\simeq\epsilon_{1,\uin}\left( \frac{V_\uin}{V}\right)^2\, .
\eea
Therefore, $\epsilon_1$ increases if the field follows the gradient of its potential and decreases otherwise. Whether or not this increase can stop inflation in the former case depends on the potential. If the relative variations of the potential are bounded this may never happen if $\epsilon_1$ has a sufficiently small value initially. 

One can also use \Eq{eq:delta:sol} to compute the number of \efolds~spent in the USR regime. Since $\dd N/\dd\phi = H/\dot{\phi} = -3H^2\delta/V'$ where we have used the definition~(\ref{eq:def:f}), in the quasi de-Sitter limit where $H^2\simeq V/(3\Mp^2)$, one obtains $\dd N/\dd \phi = -V\delta/(V'\Mp^2)$. Making use of \Eq{eq:delta:sol}, this gives rise to $\dd N/\dd \phi = - V \delta_\uin/(V'_\uin \Mp^2)$, and hence
\bea
\label{eq:DeltaN:USR}
\Delta N_\USR = -\frac{\delta_\uin}{\Mp^2V'_\uin}\int_{\phi_\uin}^\phi V(\tilde{\phi})\dd\tilde{\phi}\, .
\eea
This should be compared with the slow-roll formula $\Delta N_\SR = - 1/\Mp^2\int_{\phi_\uin}^\phi V(\tilde{\phi})/V'(\tilde{\phi})\dd\tilde{\phi} $, which shows that in general fewer \efolds~are realised between two given field values in the USR regime than in standard slow roll. 
From this slow-roll formula it is even clear that $\Delta N_\SR$ can become infinite if there is a flat point in the potential such that $V/V'$ is not integrable as $V' \to 0$. 
However, the USR formula \eqref{eq:DeltaN:USR} is always integrable and finite, even when one crosses a flat inflection point of the potential.

Note that when we find USR to be a local attractor ($|\delta|$ decreases), it is so for sufficiently small values of $|\delta|<\vert \eta_V \vert$ only. If $|\delta| > \vert \eta_V \vert$ initially, then the first term on the right-hand side of \Eq{eq:f:dynamical:USR:new} dominates even if $\epsilon_V<\eta_V^2$ and the analysis of section~\Sec{sec:stability:USR:inflation:case1} shows that USR becomes unstable. In this case trajectories diverge from USR ($f\sim 1$) to approach the standard slow roll ($f\ll1$) for $\epsilon_V\ll1$ and $|\eta_V|\ll1$. This shows that, if $\epsilon_V<\eta_V^2 \ll 1$, the boundary between the SR and the USR basins of attraction is located around the line $\vert\delta\vert\sim\vert\eta_V\vert$. This will be checked explicitly in the example presented in \Sec{sec:example:inflectionPoint}.
This is similar to bifurcation behaviour that has previously been discussed for inflection point quintessence \cite{Chang:2013cba}.

In summary, we find that if the inflaton rolls down its potential, USR inflation is stable if $V''>0$ and $\eta_V^2>\epsilon_V$, which can be combined into the condition
\bea
\label{eq:USR:stable:criterion}
\eta_V>\sqrt{\epsilon_V}\, ,
\eea
and continues to inflate provided $V/V_\uin$ remains larger than $\sqrt{\epsilon_{1,\uin}}$. 

\section{Examples}
\label{sec:examples}

Let us now illustrate the stability analysis performed in the previous section with two examples. In the first one, the potential has a discontinuity in its slope which produces a transient regime of USR inflation. In the second one, the potential has a flat infection point around which the inflaton field evolves in the USR regime.

\subsection{Starobinsky inflation}
\label{sec:Staro}
\begin{figure}[t]
\begin{center}
\includegraphics[width=0.65\textwidth]{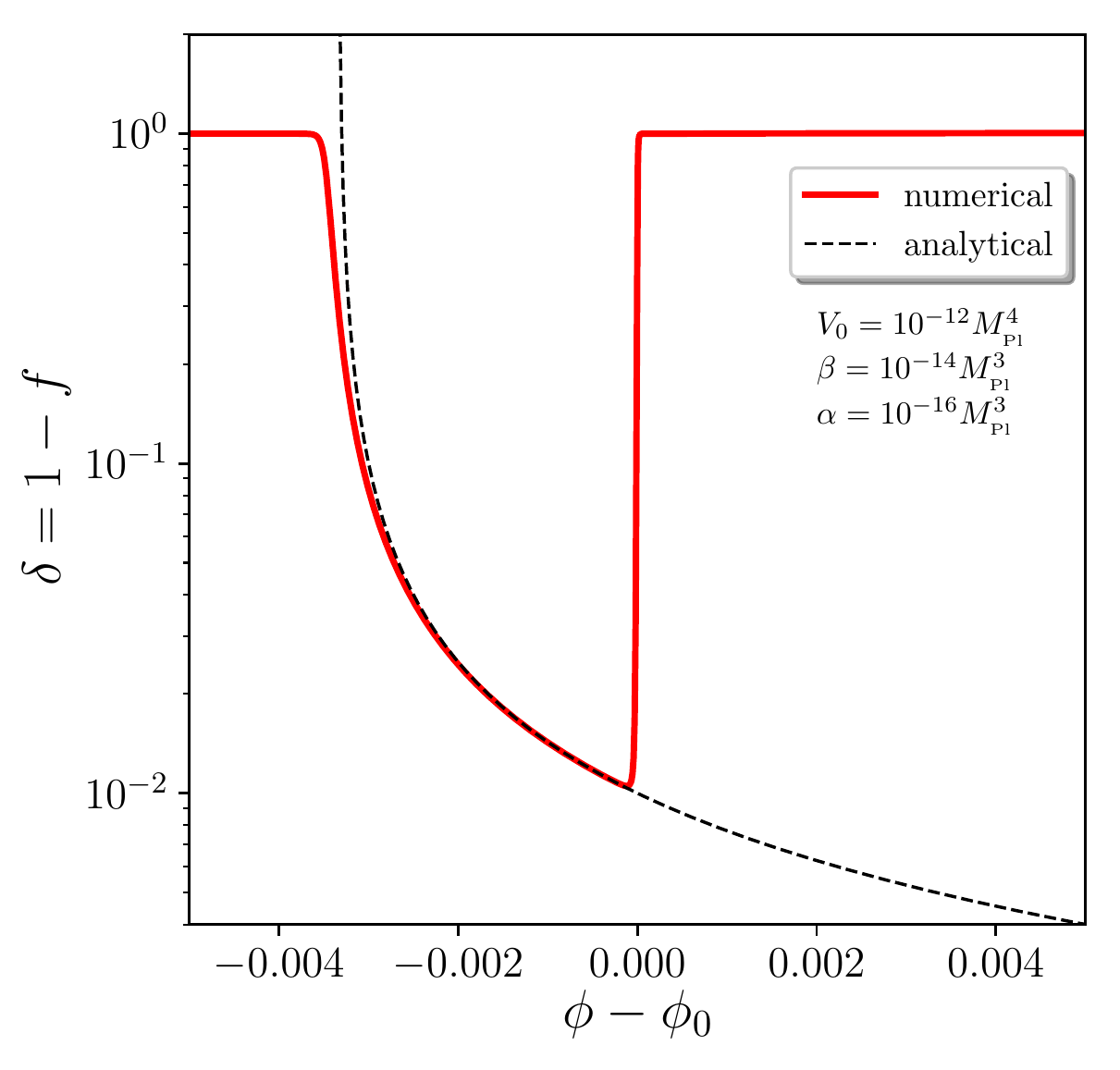}
\caption{Field acceleration parameter $\delta=1-f$ in the Starobinsky model~(\ref{eq:def:pot:strobinsky}) as a function of the field value. Before crossing the discontinuity point, a regime of SR inflation takes place where $\delta\simeq 1$. Right after crossing $\phi=\phi_0$, $\delta$ drops to small values which signals the onset of a USR phase of inflation, that quickly transitions towards a new SR phase. The solid red curve is obtained from numerically solving \Eqs{eq:eom:kleingordon} and~(\ref{eq:friedmann}) and making use of \Eq{eq:def:f}, while the black dashed curve corresponds to the analytical approximation~(\ref{eq:Staro:delta:appr}). One can check that it provides a good fit to the numerical result when $\vert\delta\vert\ll 1$.}
\label{fig:Staro}
\end{center}
\end{figure}
Let us first analyse the Starobinsky model~\cite{Starobinsky:1992ts}, where the potential is made up of two linear segments with different gradients,
\bea
\label{eq:def:pot:strobinsky}
V(\phi) = \begin{cases}
V_0 + \alpha \left( \phi-\phi_0 \right) &\text{for $\phi < \phi_0$}\\
V_0 + \beta \left( \phi-\phi_0 \right) &\text{for $\phi > \phi_0$} 
\end{cases}
\, ,
\eea
where $\beta>\alpha>0$.

Starting with $\phi>\phi_0$, the inflaton quickly relaxes to the slow-roll attractor for $V_0\gg\beta \Mp$ (corresponding to $\epsilon_V\ll1$), where, according to \Eq{eq:slowroll}, $3H\dot{\phi}\simeq-\beta$. Right after crossing $\phi=\phi_0$ where the gradient of the potential is discontinuous, $\dot{\phi}$ is still given by the same value (since the equation of motion~(\ref{eq:eom:kleingordon}) for $\phi$ is second order, $\dot{\phi}$ is continuous through the discontinuity point) but the value of $V'$ is now different, such that $f$ given by \Eq{eq:def:f} reads
\bea
\label{eq:Staro:fminus}
f_{-}= 1+\frac{V'_-}{3(H\dot{\phi})_-} = 1+\frac{V'_-}{3(H\dot{\phi})_+}\simeq 1-\frac{V'_-}{V'_+}
= 1-\frac{\alpha}{\beta}
\, .
\eea
In this expression, a subscript ``$-$'' (or ``$+$'') means that the quantity is evaluated at $\phi\rightarrow \phi_0$ with $\phi<\phi_0$ (or $\phi>\phi_0$, respectively). If $\alpha\ll \beta$, $f_-\simeq 1$ and a phase of USR is triggered. 

The analysis of \Sec{sec:stability:USR} revealed that the stability of USR inflation depends on whether $\epsilon_V$ is smaller or larger than $\eta_V^2$. In the present model, since $\eta_V$ exactly vanishes, one necessarily falls in the later case, \ie the case discussed in \Sec{sec:stability:USR:inflation:case1} where it was shown that USR inflation is always unstable. Let us also notice that in the Starobinsky model, \Eq{eq:f:dynamical:USR:case1} can be integrated analytically, and making use of \Eq{eq:Staro:fminus} for the initial condition, one finds
\bea
\label{eq:Staro:delta:appr}
\delta\simeq \frac{\alpha}{\beta+\frac{3V_0}{\Mp^2}\left(\phi-\phi_0\right)}
\, .
\eea
Since $\phi$ decreases as a function of time, $\delta$ increases, and this confirms that USR is unstable in the Starobinsky model.

These considerations are numerically checked in \Fig{fig:Staro}. One can see that when the inflaton field crosses the discontinuity point at $\phi=\phi_0$, a phase of USR inflation with small values of $\delta$ starts, which \Eq{eq:Staro:delta:appr} accurately describes. This regime is however unstable and when the inflaton field crosses the value
\bea
\phi_{\USR\rightarrow\SR}=\phi_0-\frac{\Mp^2\left(\beta-\alpha\right)}{3V_0}
\, ,
\eea 
$\delta\simeq 1$ and the system relaxes back to SR. Making use of \Eq{eq:USR:traj}, one can also estimate the number of \efolds~spent in the USR regime between the field values $\phi_0$ and $\phi_{\USR\rightarrow\SR}$, and one finds
\bea
N_\USR\simeq\frac{1}{3}\ln\left(\frac{\beta}{\alpha}\right)\, .
\eea
The number of USR \efolds~is therefore of order a few or less in this model.
\subsection{Cubic inflection point potential}
\label{sec:example:inflectionPoint}
Let us now consider the case where the potential contains a flat inflection point at $\phi=0$, around which it can be expanded as
\bea 
\label{eq:pot:inflectionPoint:cubic}
V(\phi) = V_0 \left[1+\left(\frac{\phi}{\phi_0}\right)^3\right]\, .
\eea
One could parametrise the potential with a higher odd power of the field, say $V\propto 1+(\phi/\phi_0)^5$, but this would not change the qualitative conclusions that we draw below. The potential~(\ref{eq:pot:inflectionPoint:cubic}) has a flat inflection point at $\phi=0$, where $V'=V''=0$. 
In the slow-roll regime, it takes the inflaton an infinitely long time to reach the inflection point, which it never crosses.
However in the USR regime the inflaton can traverse the inflection point in a finite time, which we estimate below. 

As explained in \Eq{eq:sr:consistency}, SR inflation requires $\epsilon_V\ll 1$ and $\vert\eta_V\vert\ll 1$, where the potential slow-roll parameters \eqref{eq:epsV} are here given by
\bea
\label{eq:InflectionPoint:SRpotParam}
\epsilon_V = \frac92 \frac{\Mp^2}{\phi_0^2} \frac{(\phi/\phi_0)^4}{\left[ 1+(\phi/\phi_0)^3 \right]^2} \,, \\
\eta_V = 6 \frac{\Mp^2}{\phi_0^2} \frac{\phi/\phi_0}{1+(\phi/\phi_0)^3} \,. 
\eea
Let us first focus on the part of the potential located before the inflection point, \ie at $\phi>0$. The parameter $\epsilon_V$ vanishes at $\phi=0$ and at $\phi\rightarrow\infty$, and in between it reaches a maximum at $\phi=2^{1/3}\phi_0$ where its value is $\epsilon_{V,\umax}=2^{1/3}\Mp^2/\phi_0^2$. The parameter $\eta_V$ has a similar behaviour, with a maximum at $\phi=2^{-1/3}\phi_0$ where its value is $\eta_{V,\umax}=2^{5/3}\Mp^2/\phi_0^2$. 

Two regimes need therefore to be distinguished: (i) if $\phi_0\gg \Mp$, SR inflation can be realised for all $\phi>0$, while (ii) if $\phi_0\ll \Mp$, SR inflation only takes place at sufficiently large ($\phi\gg\Mp$) or sufficiently small ($\phi\ll \phi_0^3/\Mp^2$) field values. After crossing the inflection point at  $\phi=0$, the potential decreases towards zero and the potential slow-roll parameter, $\epsilon_V$, diverges, signalling the end of inflation, so we restrict our analysis to the field values $\phi>-\phi_0$.

USR inflation can be studied making use of the results of \Sec{sec:stability:USR}, where it was shown that USR inflation is stable if $\eta_V>\sqrt{\epsilon_V}$, see \Eq{eq:USR:stable:criterion}. Together with \Eq{eq:InflectionPoint:SRpotParam}, this gives rise to the USR stability condition
\bea
\label{eq:InflectionPoint:USR:stability:Condition}
0<\phi<2\sqrt{2}\Mp\, .
\eea

We shall now study the two regimes $\phi_0\gg\Mp$ and $\phi_0\ll\Mp$ separately.

\subsubsection{Case $\phi_0\gg \Mp$}
\begin{figure}
\begin{center}
\includegraphics[width=0.45\textwidth, height=6cm]{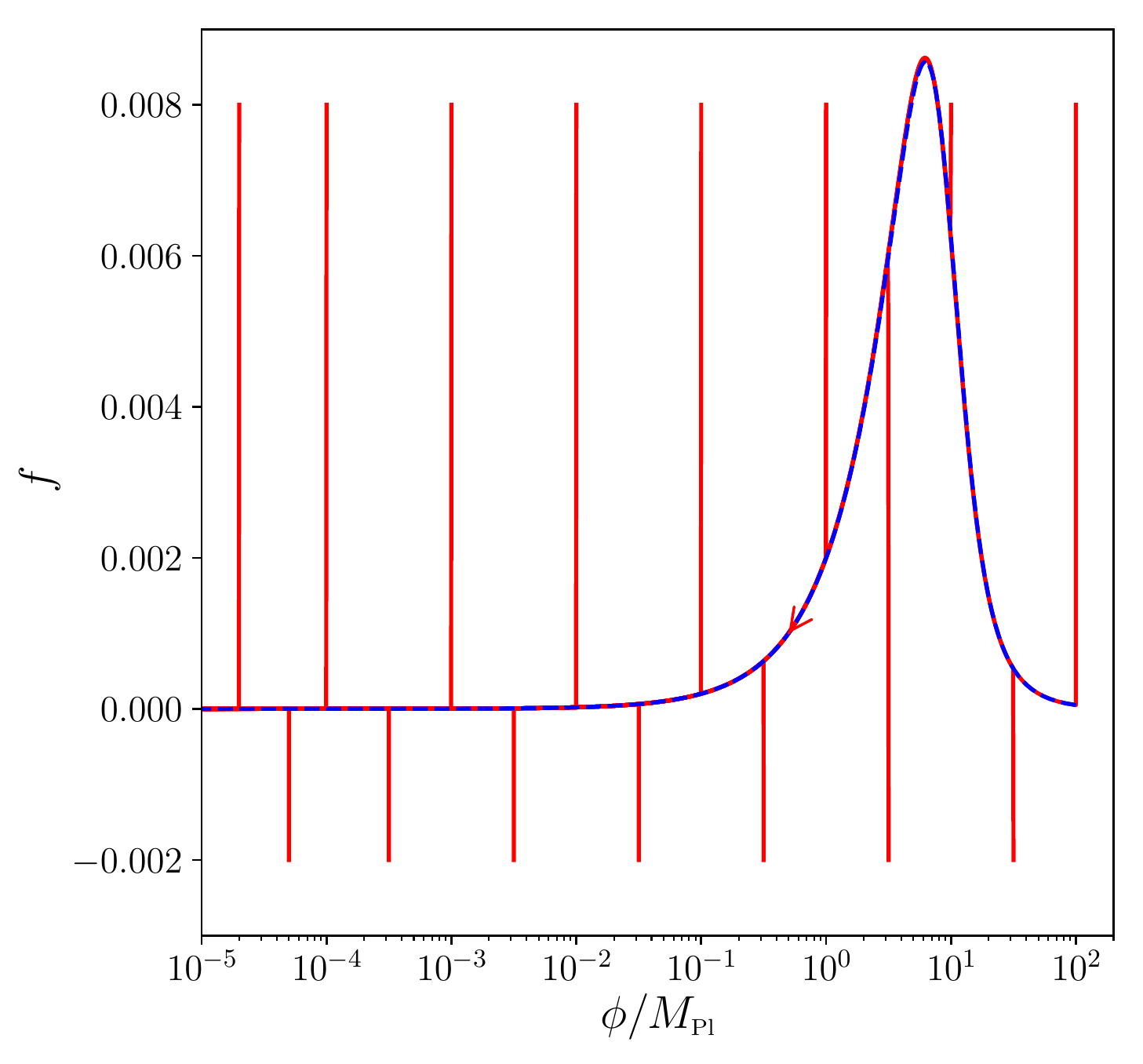} 
\includegraphics[width=0.45\textwidth, height=6.1cm]{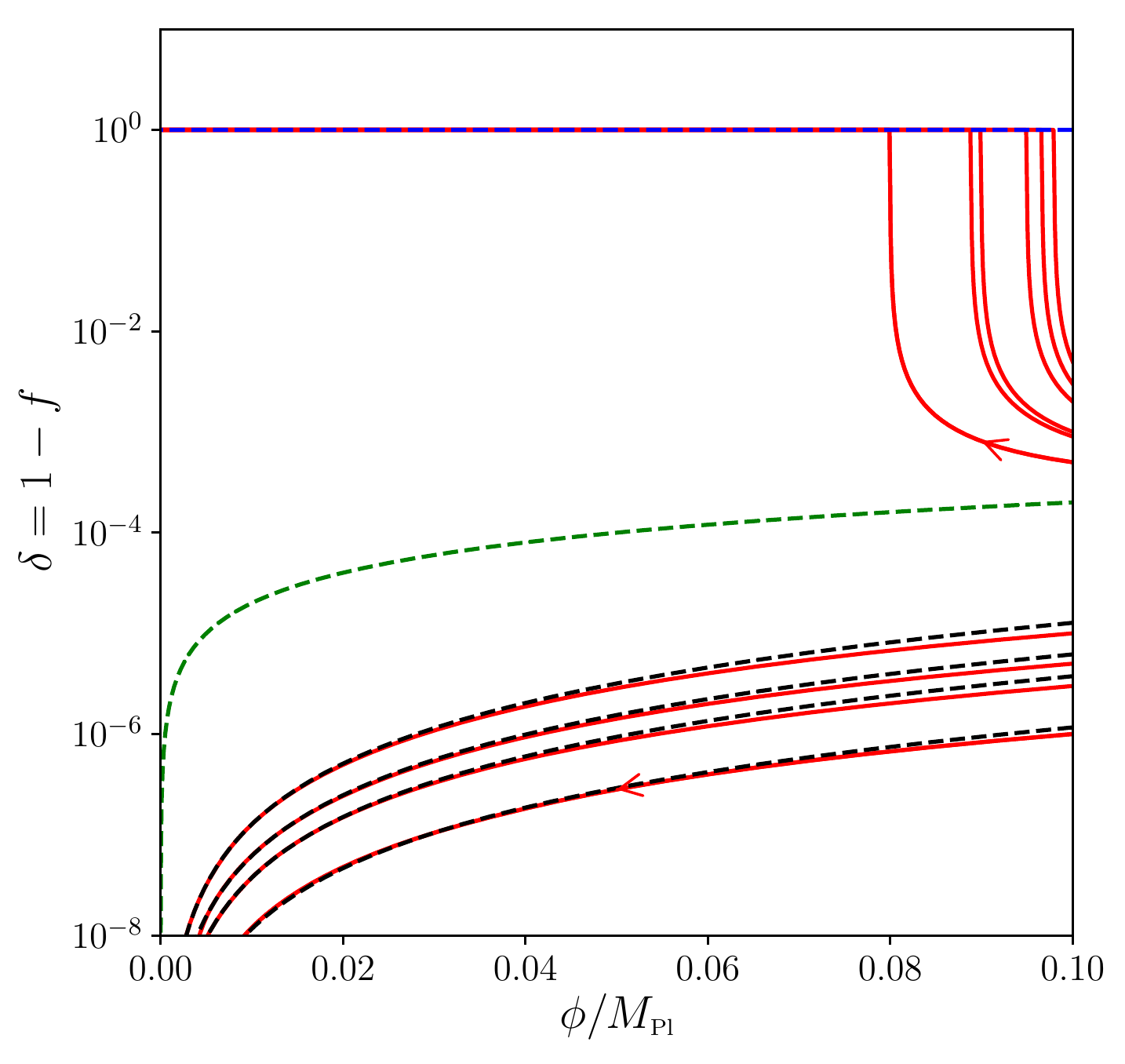} 
\caption{Field acceleration parameter in the cubic inflection point model~(\ref{eq:pot:inflectionPoint:cubic}) as a function of the field value, for $\phi_0=10\Mp$ and $V_0=4.2\times 10^{-11}$. The red lines correspond to numerical solutions of \Eq{eq:f:dynamical} and the dashed blue line stands for the slow-roll limit~(\ref{eq:f:SR}). The left panel zooms in on the region $f\simeq 0$ where one can see that SR is an attractor. The right panel uses a logarithmic scale on $1-f=\delta$, such that it zooms in on the USR regime $f\simeq 1$. If $\delta$ is initially smaller than $\vert \eta_V \vert/3$, represented with the dashed green line, the trajectories evolve towards $\delta=0$, otherwise they evolve to reach the SR attractor. The black dotted lines correspond to the analytical USR approximation~(\ref{eq:delta:sol}).
}
\label{fig:InflectionPoint:fsol:phi0GTMp}
\end{center}
\end{figure}
In this case, as already mentioned, SR inflation is an attractor over the entire range $\phi>0$ (until inflation stops when $\phi$ approaches $-\phi_0$). This implies that if one starts from an initial field value that is larger than the USR stability upper bound given in \Eq{eq:InflectionPoint:USR:stability:Condition}, $\phi=2\sqrt{2}\Mp$, the system relaxes towards the SR attractor (SR is the only stable attractor at $\phi>2\sqrt{2}\Mp$) and stays in SR until the end of inflation. In this scenario, even though USR inflation is also a local attractor at $\phi<2\sqrt{2}\Mp$, the inflaton field never drives a phase of USR inflation. 

The only way to get a period of USR inflation is therefore to start with $\phi<2\sqrt{2}\Mp$. There, as explained in \Sec{sec:stability:USR}, USR inflation is stable and its basin of attraction is bounded by the condition $\sqrt{\epsilon_V}<|\delta|<\eta_V$.

These considerations are numerically verified in \Fig{fig:InflectionPoint:fsol:phi0GTMp}. In the left panel, the SR region $\vert f\vert\ll 1 $ is displayed, where one can check that the numerical solutions of \Eq{eq:f:dynamical} (red curves) all converge towards the SR attractor~(\ref{eq:f:SR}) (dashed blue curve). In the right panel, a logarithmic scale is used on $1-f$, which allows one to zoom in on the USR region $f\simeq 1$. Since the initial values for $\phi$ satisfy \Eq{eq:InflectionPoint:USR:stability:Condition}, one can check that the trajectories with $\delta<\eta_V$ converge towards USR, while the ones for which $\delta>\eta_V$ approach the SR attractor. This confirms that the boundary between the two basins of attraction is located around the line $|\delta|=|\eta_V|$. The analytical approximation~(\ref{eq:delta:sol}) is displayed with the black dotted lines and one can check that it provides a good fit to the numerical result in the USR regime.

Let us finally estimate the number of \efolds~that is typically realised in the USR inflating regime. In the stability range~(\ref{eq:InflectionPoint:USR:stability:Condition}) of USR inflation, the potential is dominated by its constant piece since $\phi_0\gg\Mp$. The first Hubble-flow parameter is therefore roughly constant during the USR epoch, see \Eq{eq:USR:stable:eps1:appr}. Starting USR inflation at $\phi_\uin\sim \Mp$ with $\delta=\delta_\uin$, \Eq{eq:delta:sol} implies that $\delta$ goes back to its initial value $\delta_\uin$ at around $\phi\sim -\Mp$.  Plugging these values into \Eq{eq:DeltaN:USR}, one obtains
\bea
\label{eq:FlatInflectionPoint:NUSR:phi0GTMp}
\Delta N_\USR\simeq \frac{2\delta_\uin}{3}\left(\frac{\phi_0}{\Mp}\right)^3.
\eea
This shows that, in the regime $\phi_0\gg \Mp$, a large number of USR~\efolds~can be realised.
However, we should note that this number remains finite, contrary to what happens in the slow-roll regime where it takes an infinite time to cross the inflection point, as already mentioned.

\subsubsection{Stochastic diffusion}
This large number of USR inflationary \efolds~is however derived under the assumption that the field behaves classically all the way down to the inflection point, while stochastic diffusion is expected to play a role when the potential becomes very flat. Let us estimate how this changes the above result.

Starting from $\phi_\uin=\Mp$ and $\delta=\delta_\uin$ as explained above, \Eq{eq:delta:sol} leads to $\delta(\phi)\simeq \delta_\uin(\phi/\Mp)^2$ (where we assume $\phi<\Mp$). Then, making use of \Eq{eq:DeltaN:USR}, if the field behaved in a purely classical manner, the number of \efolds~realised between $\phi$ and $-\phi$ would be given by $\Delta N_\USR(\phi)\simeq 2\delta_\uin/3 (\phi_0/\Mp)^3(\phi/\Mp)$.

On the other hand, if the field was only driven by stochastic noise, its equation of motion would be given by~\cite{Starobinsky:1986fx} $\dd\phi/\dd N=H/(2\pi)\xi$, where $\xi$ is a white Gaussian noise with vanishing mean and unit variance, such that $\langle \xi(N) \xi(N') \rangle = \delta(N-N')$.  Assuming that $H$ is roughly constant, this leads to $\langle \phi^2 \rangle = H^2/(2\pi)^2 N$, hence the typical number of \efolds~required for the inflaton field value to go from $\phi$ to $-\phi$ is given by $\Delta N_{\mathrm{sto}}= 48\pi^2\phi^2\Mp^2/V_0$. Notice that this can also be obtained using the ``first-passage-time techniques'' developed in \Ref{Vennin:2015hra}. Setting a reflective boundary condition at $\phi$ and an absorbing one at $-\phi$, one finds that the mean number of \efolds~required to reach $-\phi$ starting from $\phi$ exactly coincides with the expression we just wrote for  $\Delta N_{\mathrm{sto}}$.

Since $\Delta N_\USR$ scales as $\phi$ and $\Delta N_{\mathrm{sto}}$ as $\phi^2$, two regimes need to be distinguished. When $\phi>\phi_{\mathrm{sto}}$, where 
\bea
\frac{\phi_{\mathrm{sto}}}{\phi_0} = \frac{\delta_\uin}{72\pi^2}\left(\frac{\phi_0}{\Mp}\right)^2\frac{V_0}{\Mp^4}
\eea
is the solution of $\Delta N_\USR(\phi_{\mathrm{sto}})=\Delta N_{\mathrm{sto}}(\phi_{\mathrm{sto}})$, one has $\Delta N_\USR<\Delta N_{\mathrm{sto}}$, which means that classical USR is more efficient at driving the field than stochastic diffusion, hence the dynamics of the field are essentially classical. When $\phi<\phi_{\mathrm{sto}}$ on the other hand, stochastic diffusion takes over, which means that the part of the potential where $-\phi_{\mathrm{sto}}<\phi<\phi_{\mathrm{sto}}$ is dominated by quantum diffusion.  

This is why, for classical USR to take place, one needs to impose $\phi_{\mathrm{sto}}<\Mp$, which means that the potential energy cannot be too large,
\bea
\frac{V_0}{\Mp^4}\ll 72\pi^2\left(\frac{\Mp}{\phi_0}\right)^3
\eea
(recall that $\phi_0\gg \Mp$ so this is not necessarily guaranteed). When this is the case, the number of \emph{classical} USR infationary \efolds~is given by $\Delta N_{\USR {}_{,\mathrm{class}}} = \Delta N_{\USR}(\Mp)-\Delta N_{\USR}(\phi_{\mathrm{sto}} )$, where $\Delta N_{\USR}(\Mp)$ was given in \Eq{eq:FlatInflectionPoint:NUSR:phi0GTMp}, and one obtains
\bea
\Delta N_{\USR{}_{,\mathrm{class}}} = \frac{2\delta_\uin}{3}\left(\frac{\phi_0}{\Mp}\right)^3 \left[1 - \frac{\delta_\uin}{72\pi^2}\left(\frac{\phi_0}{\Mp}\right)^3 \frac{V_0}{\Mp^4}\right]\, .
\eea
If the parameter $\phi_0$ is chosen such that $1\ll \phi_0/\Mp \ll (72\pi^2V_0/\Mp^4)^{-1/3}$, this number can still be very large and a sustained phase of classical USR inflation takes place.
\subsubsection{Case $\phi_0\ll \Mp$}
\begin{figure}
\begin{center}
\includegraphics[width=0.45\textwidth, height=6cm]{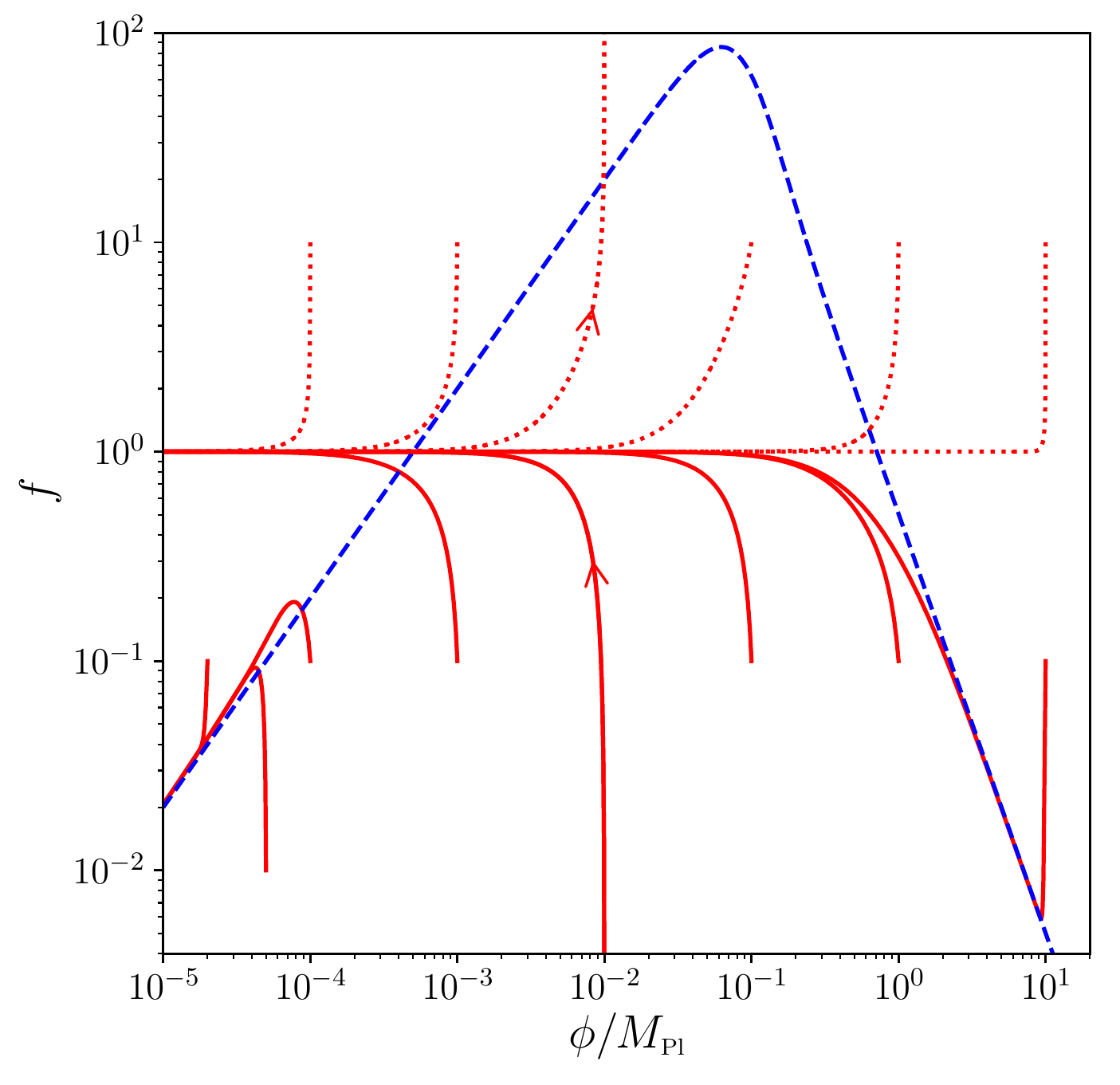} 
\includegraphics[width=0.45\textwidth, height=6.1cm]{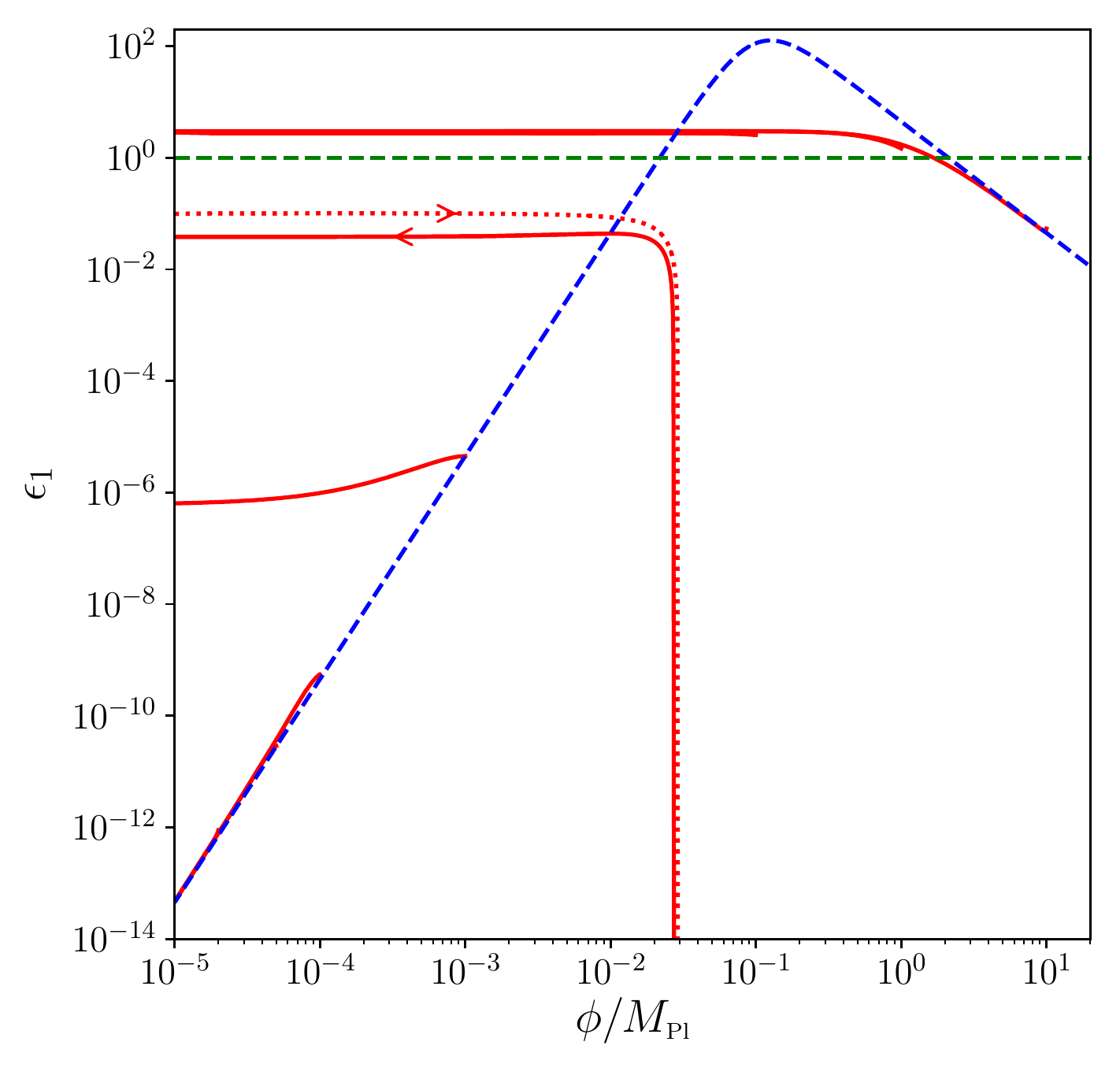} 
\caption{Left panel: Field acceleration parameter in the cubic inflection point model~(\ref{eq:pot:inflectionPoint:cubic}) as a function of the field value, for $\phi_0=0.1\Mp$ and $V_0=4.2\times 10^{-11}$, with the same conventions as in \Fig{fig:InflectionPoint:fsol:phi0GTMp}. The solid part of the red curves correspond to when $f<1$ and $\phi$ decreases with time, while the dotted parts are for $f>1$ and $\phi$ increases (as indicated by the arrows). Right panel: first Hubble-flow parameter $\epsilon_1$ as a function of the field value for the same solid trajectories and the dotted trajectory with an arrow in the left panel. The dashed green line stands for $\epsilon_1=1$ below which inflation proceeds. The trajectories that have both $\epsilon_1 \ll 1$ and $f \to 1$ drive a phase of USR inflation. 
}
\label{fig:InflectionPoint:fsol:phi0LTMp}
\end{center}
\end{figure}
In this case, SR inflation can only occur at $\phi\gg\Mp$ or $\phi\ll\phi_0^3/\Mp^2$. One therefore has three regions: if $\phi\gg\Mp$, SR is the only attractor, if $\phi_0^3/\Mp^2\ll\phi\ll\Mp$, USR is the only attractor, and if $\phi\ll \phi_0^3/\Mp^2$, both SR and USR are local attractors. These three regimes can be clearly seen in the left panel of \Fig{fig:InflectionPoint:fsol:phi0LTMp}, where the same colour code as in \Fig{fig:InflectionPoint:fsol:phi0GTMp} is employed. In the right panel of \Fig{fig:InflectionPoint:fsol:phi0LTMp}, the first Hubble-flow parameter is displayed for the same trajectories. The solid curves have $f<1$ for which $\phi$ decreases with time and the dotted curves have $f>1$ for which $\phi$ increases with time. We shall now discuss each of these three regimes in more detail.

Firstly, if one starts with an initial value of $\phi$ that is super Planckian, one quickly reaches the SR attractor. Then when $\phi$ becomes of order $\Mp$, $f_\SR$ becomes of order one which signals the breakdown of SR and one leaves the SR line to settle down to $f\simeq 1$, \ie in the USR regime. However, as can be seen on the right panel of \Fig{fig:InflectionPoint:fsol:phi0LTMp}, the first Hubble-flow parameter converges towards $\epsilon_1\simeq 3$, so inflation stops around $\phi\simeq\Mp$ and does not resume afterwards. In this case, for $\phi<\Mp$, we have USR but not USR inflation, and this non-inflating USR regime is stable due to the considerations of footnote~\ref{footnote:USR:Non:Inflation}.

Secondly, if one starts with an initial field value between $\phi_0^3/\Mp^2$ and $\Mp$ and with $\dot\phi<0$ (rolling down the potential), the field converges towards USR, since it is the only stable solution. This is the case for the trajectory with $f<1$ on which an arrow has been added in \Fig{fig:InflectionPoint:fsol:phi0LTMp}. Let us recall that the dotted part of the trajectory corresponds to $f>1$ and the inflaton climbs up its potential ($\dot\phi>0$), until its velocity changes vanishes at which point $f$ diverges and $\dot\phi$ changes sign. The inflaton then rolls down its potential starting from very negative values for $f$ (solid part of the curve) and quickly reaches USR. On the right panel of \Fig{fig:InflectionPoint:fsol:phi0LTMp} one can see that $\epsilon_1$ is roughly constant in the rolling down phase, which is consistent with \Eq{eq:USR:stable:eps1:appr} since the potential is dominated by its constant piece $V\simeq V_0$ when $\phi\ll\phi_0$.
\begin{figure}
\begin{center}
\includegraphics[width=0.65\textwidth]{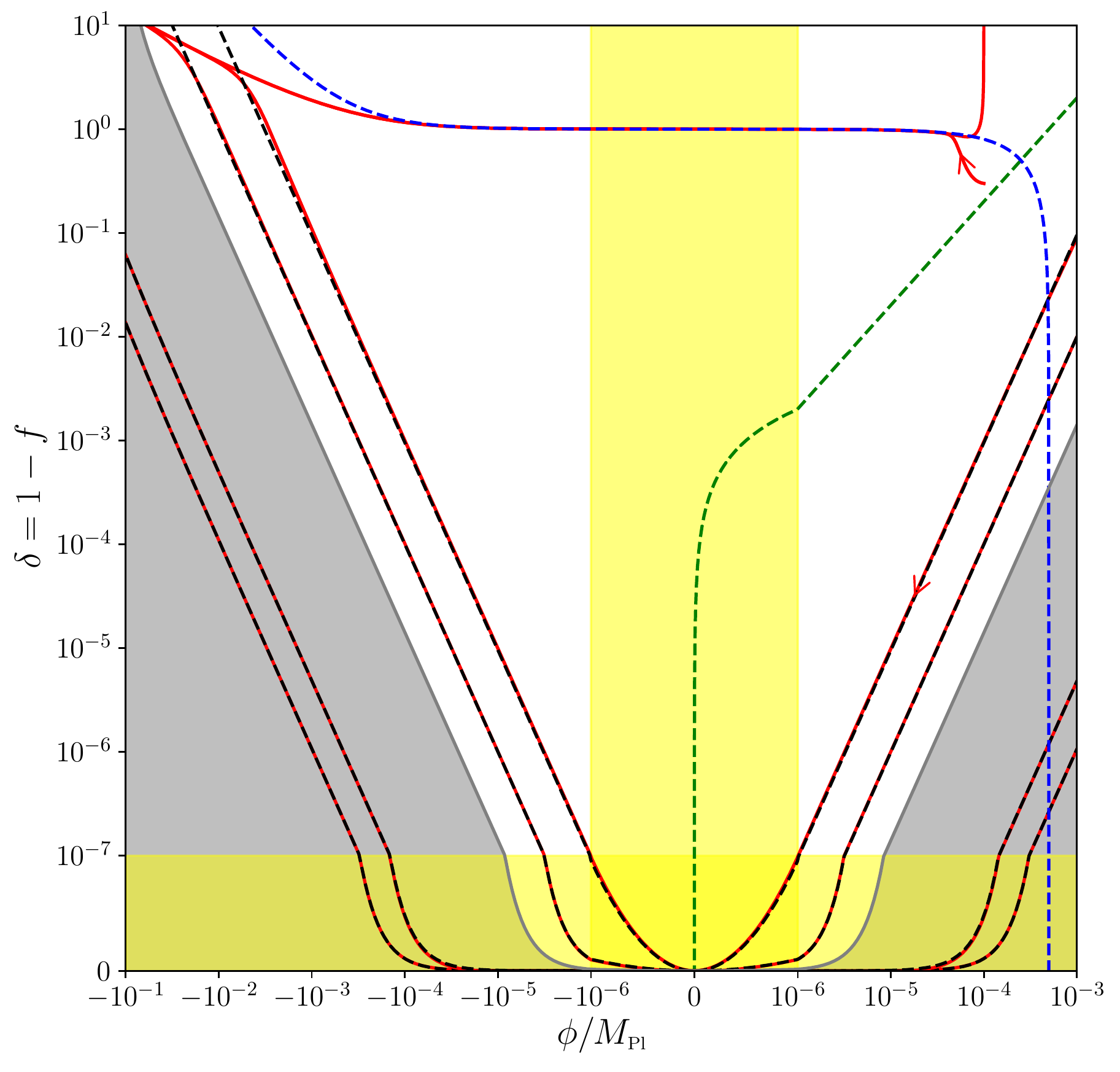} 
\caption{Same as in the left panel of \Fig{fig:InflectionPoint:fsol:phi0LTMp} for the region $-\phi_0<\phi<\phi_0^3/\Mp^2$. Yellow shading denotes regions when the axis scale is linear rather than logarithmic, and the grey shaded region is where inflation in not happening $\epsilon_1>1$, as given by \eqref{eq:inflation:condition}. The black dashed lines stand for the analytical USR inflation approximation~(\ref{eq:delta:sol}) in the inflating part, and to the USR \emph{non}-inflation approximation~(\ref{eq:delta:sol:USRnoninflating}) in the non-inflating part (the field excursion being sub Planckian, the two behaviours are very much similar). The dashed green line stands for $\delta=\vert\eta_V\vert/3$, which in the $\phi>0$ region corresponds to the boundary between the SR and the USR basins of attraction. For $\phi<0$, only SR is an attractor which explains why those trajectories that reach the USR attractor in the $\phi>0$ region have $\delta$ increasing with time in the $\phi<0$ region where USR is unstable. 
}
\label{fig:InflectionPoint:fsol:phi0LTMp:phiLTphi0power3}
\end{center}
\end{figure}

Lastly, if one starts with $\phi\ll \phi_0^3/\Mp^2$, one either reaches the SR attractor if $|\delta|>|\eta_V|$ or the USR attractor if $|\delta|<|\eta_V|$. This can be more clearly seen in \Fig{fig:InflectionPoint:fsol:phi0LTMp:phiLTphi0power3}, where the whole region $-\phi_0<\phi<\phi_0^3/\Mp^2$ is displayed. One can check that the inflating USR approximation~(\ref{eq:delta:sol}) provides a good approximation to the numerical solutions of \Eq{eq:f:dynamical} in the inflating part of phase space for those trajectories that reach the USR attractor, while the \emph{non}-inflating USR approximation~(\ref{eq:delta:sol:USRnoninflating}) correctly describes the non-inflating trajectories (in the grey shaded region of the plot). Here, because the field excursion is sub-Planckian (since $\phi_0\ll\Mp$), these two behaviours are almost identical. As in the right panel of \Fig{fig:InflectionPoint:fsol:phi0GTMp}, one can also check that the line $\vert\delta\vert\sim\vert\eta_V\vert$ correctly delimitates the boundary between the two basins of attraction when $\phi>0$. 
If $\phi<0$, USR becomes unstable and only SR remains as an attractor, which explains why $\delta$ increases with time for those trajectories that reached the USR attractor before crossing the flat inflection point. However, one should note that those trajectories do not have time to reach the SR attractor before the potential becomes too steep and SR is violated.
Similarly, for $\phi > \phi_0^3/\Mp^2$ the potential is too steep and SR is not a valid approximation.

Let us also estimate the number of \efolds~that is typically realised in the USR inflating regime. USR is stable in the range~(\ref{eq:InflectionPoint:USR:stability:Condition}). However, for $\phi>\phi_0$, the potential is not dominated by its constant piece so $\epsilon_1$ can substantially increase because of \Eq{eq:USR:stable:eps1:appr}. Whether or not USR \emph{inflation} is maintained depends on the initial value of $\epsilon_1$ (see the right panel of \Fig{fig:InflectionPoint:fsol:phi0LTMp}) and to avoid this initial condition dependence, let us consider the case where we start USR inflation around $\phi\sim\phi_0$. Starting with $\delta=\delta_\uin$, \Eq{eq:delta:sol} implies that $\delta$ goes back to its initial value $\delta_\uin$ at around $\phi\sim -\phi_0$.  Making use of \Eq{eq:DeltaN:USR}, this gives rise to
\bea
\Delta N_\USR \simeq \frac{2\delta_\uin}{3}\left(\frac{\phi_0}{\Mp}\right)^2\, .
\eea
This shows that, in the regime $\phi_0\ll\Mp$, the number of \efolds~realised in the USR regime is necessarily small, contrary to the case $\phi_0\gg\Mp$, see \Eq{eq:FlatInflectionPoint:NUSR:phi0GTMp}.
\begin{figure}
\begin{center}
\includegraphics[width=0.49\textwidth]{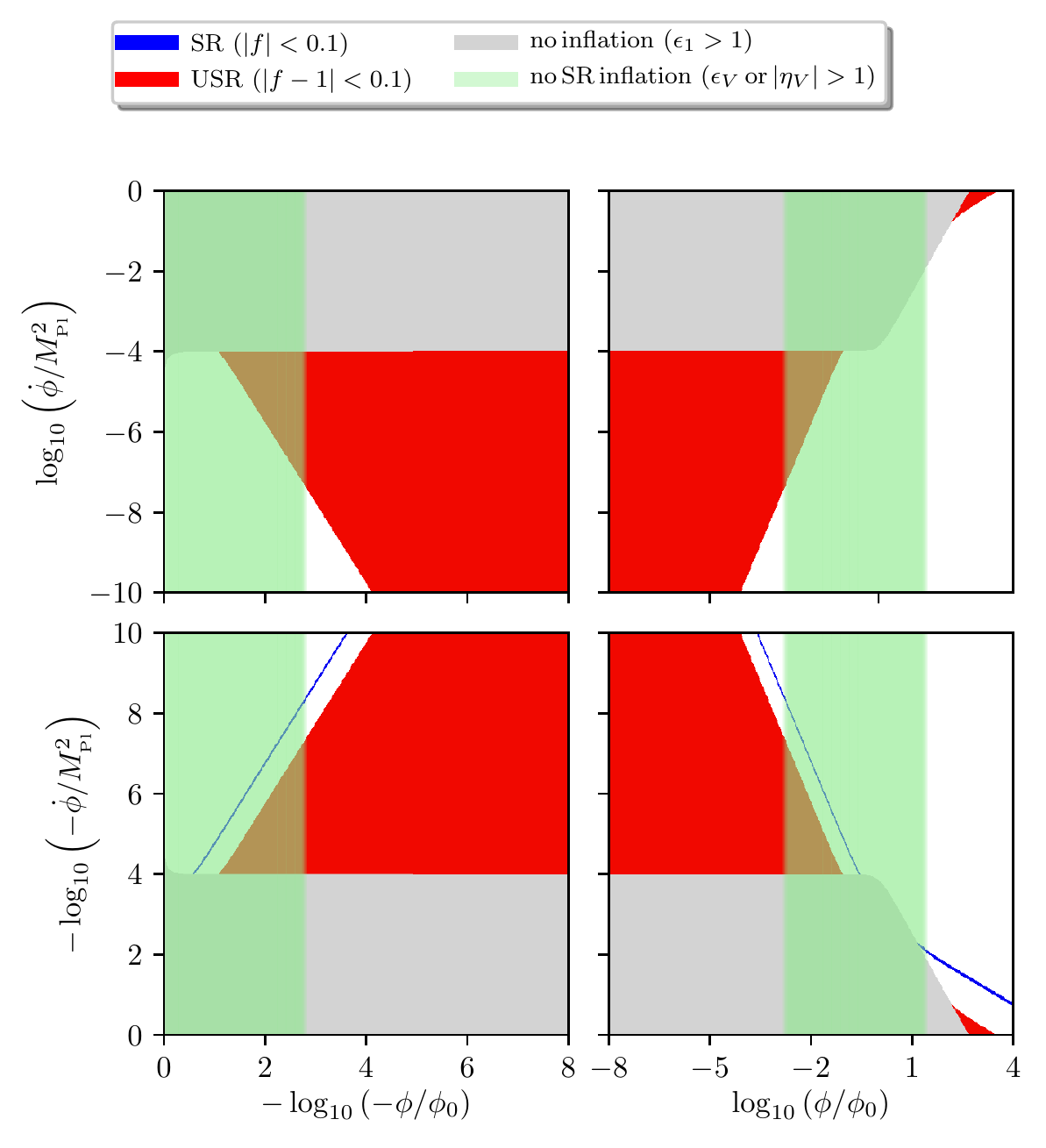} 
\includegraphics[width=0.49\textwidth]{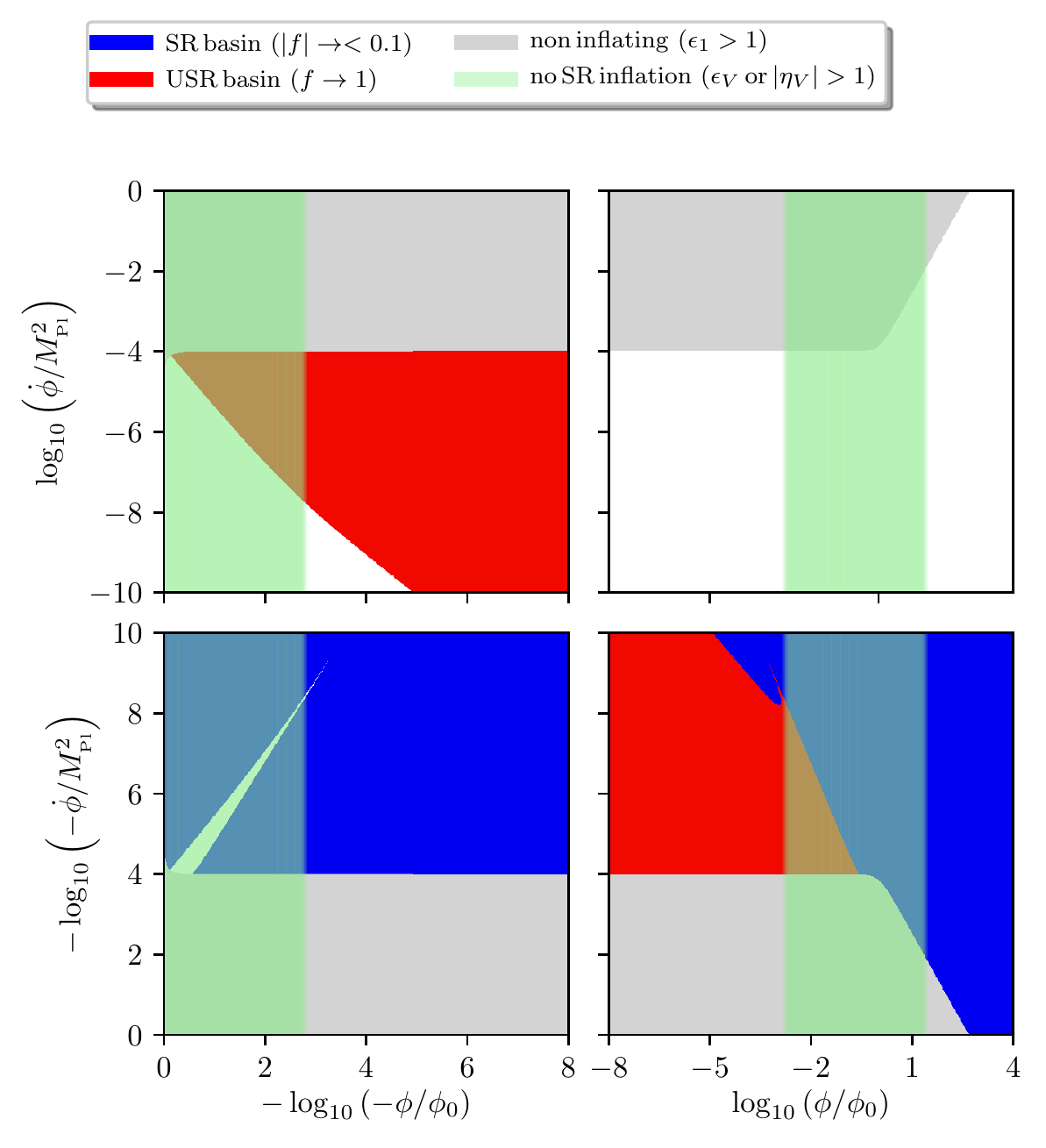} 
\caption{Regions in phase space for the cubic inflection point model~(\ref{eq:pot:inflectionPoint:cubic}) with $V_0=4.2\times 10^{-11}$ and $\phi_0=0.1\Mp$ where SR and USR solutions exist (left panel), and (right panel) basins of attraction for SR ($f<1$ and $|f|$ decreasing) and USR ($|1-f|$ decreasing).}
\label{fig:phasespace_phi0eq0p1}
\end{center}
\end{figure}

\begin{figure}
\begin{center}
\includegraphics[width=0.49\textwidth]{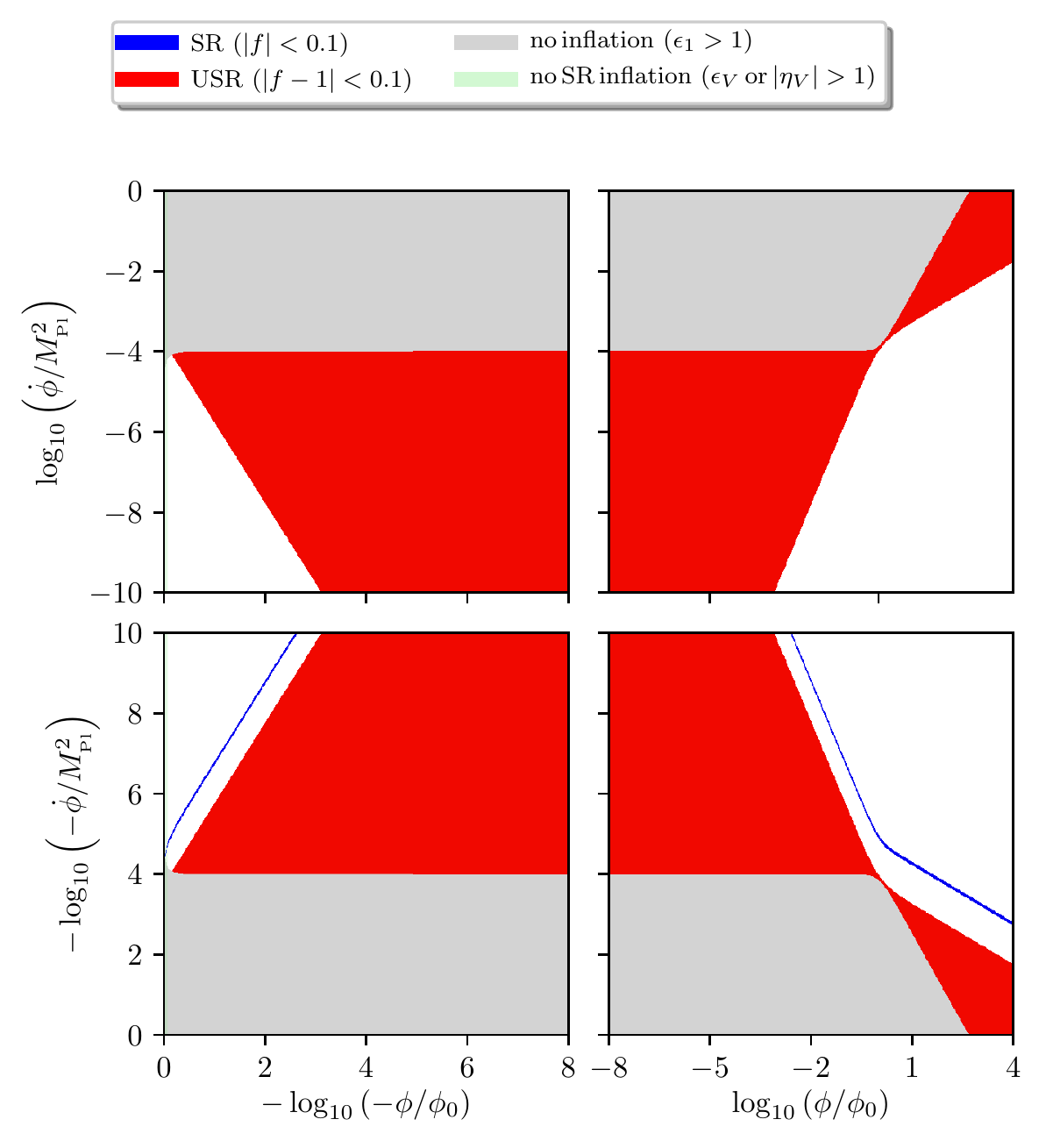} 
\includegraphics[width=0.49\textwidth]{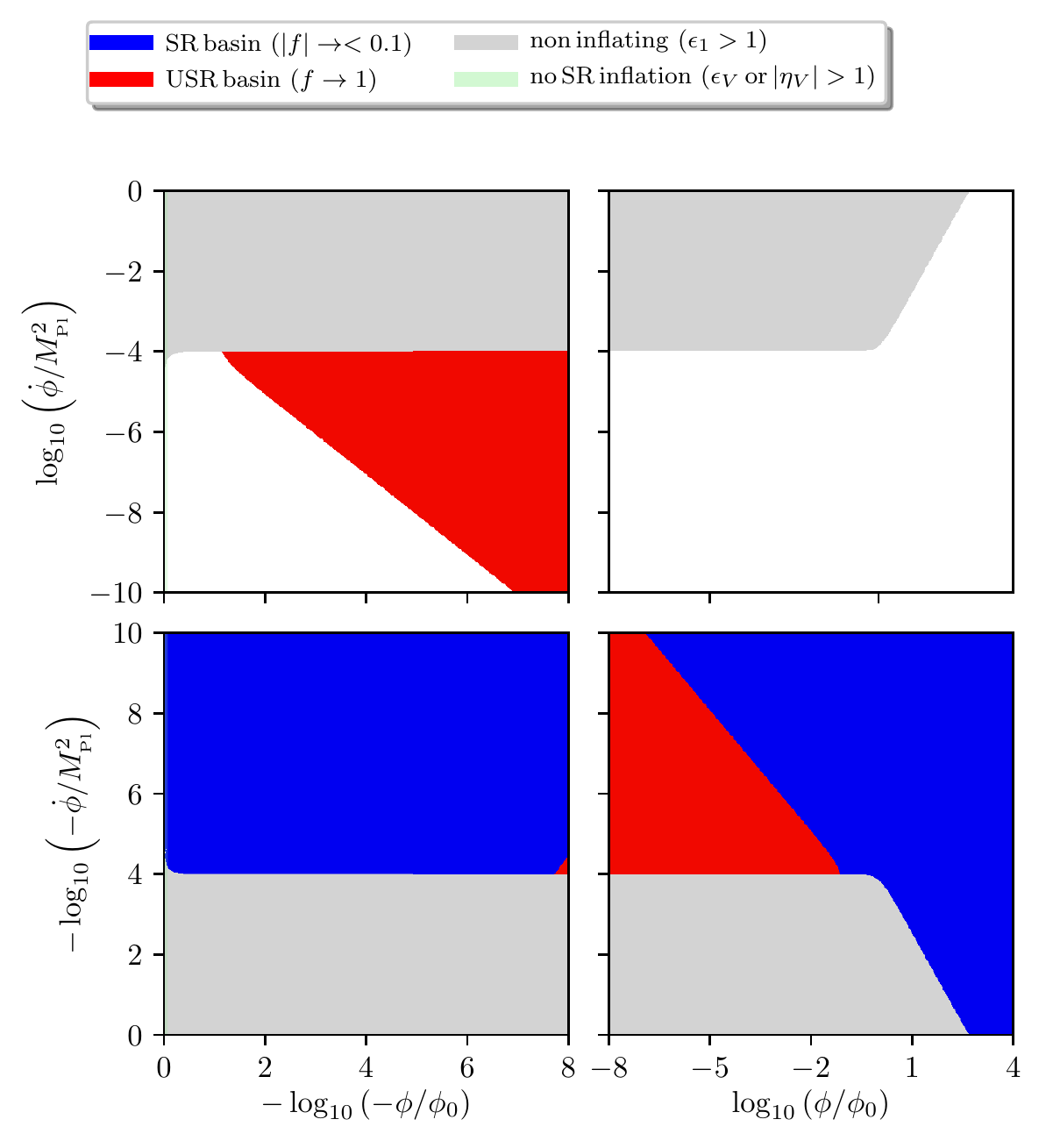} 
\caption{Regions in phase space for the cubic inflection point model~(\ref{eq:pot:inflectionPoint:cubic}) with $V_0=4.2\times 10^{-11}$ and $\phi_0=10\Mp$ where SR and USR solutions exist (left panel), and regions of stability for SR and USR (right panel).}
\label{fig:phasespace_phi0e10}
\end{center}
\end{figure}
Finally we plot in \Figs{fig:phasespace_phi0eq0p1} and~\ref{fig:phasespace_phi0e10} the phase space $(\phi,\dot\phi)$ for the cubic inflection point model~(\ref{eq:pot:inflectionPoint:cubic}), for $\phi_0=0.1\Mp$ and $\phi_0=10\Mp$ respectively. In the left panels, the blue region corresponds to SR solutions (defined as $|f|<0.1$) and the red region to USR (defined as $|1-f|<0.1$). In the right panels we show the basins of attraction of SR and USR, defined by the behaviour of $f$.

One can see that SR corresponds to a thin line in phase space while USR spans a larger region. This is due to the fact that in USR inflation, there is no unique USR trajectory in phase space and solutions retain a dependence on initial conditions as can be seen \eg in \Eq{eq:delta:sol}. This is not the case for SR that singles out a unique phase-space trajectory, see \Eq{eq:f:SR}. Note also that SR solutions only exist in the quadrants where the field velocity is aligned with the potential gradient while USR exists in every quadrant.

The right-hand plots show the basins of attraction of SR (\ie where $f<1$ and $|f|$ decreases) and USR (where $|1-f|$ decreases). When $\phi>0$, if the field goes up the potential ($\dot{\phi} >0$) then  USR is unstable, and there is no SR regime and hence no SR basin of attraction either. If the field rolls down the potential ($\dot{\phi}<0$), when $\phi\gg \Mp$ or $\phi<0$ we see that only SR is an attractor as discussed above, and when $0<\phi\ll\Mp$ both SR and USR can be attractors. When $\phi<0$ and the field goes up the potential, USR is an attractor in some region of the phase space.
This corresponds to initial conditions where the field arrives at the inflection point with an almost vanishing velocity and inflates in the USR regime.

\section{Conclusions}
\label{sec:conclusions}
In this paper we have shown that
ultra-slow-roll inflation can be a stable attractor in phase space.

%
We have performed a stability analysis in terms of the dimensionless field acceleration parameter, $f=1-\delta$ in \Eq{eq:def:f}, that quantifies the acceleration of the inflaton field relative to the Hubble friction term in the Klein-Gordon equation. 
We show that ultra-slow-roll inflation ($|\delta|\ll1$) is stable ($\dd |\delta|/ \dd t<0$) for a scalar field rolling down a convex potential ($\dot{V}<0$ and $V''>0$) if the condition~(\ref{eq:USR:stable:criterion}) is fulfilled, which in terms of the potential function $V(\phi)$ reads
\bea
\label{eq:USR:stable:criterion:conclusion}
\Mp V''>\left\vert V' \right\vert\, .
\eea
Conversely, standard slow roll ($|f|\ll1$) is always an attractor whenever the slow-roll consistency conditions \Eq{eq:sr:consistency} are satisfied.

We have compared our analytical results against numerical examples. For the Starobinsky model \eqref{eq:def:pot:strobinsky} where the potential is made of the two linear pieces, the condition~\eqref{eq:USR:stable:criterion:conclusion} is never fulfilled, since $V''=0$, and ultra-slow-roll inflation is never stable. It lasts for a number of \efolds~of order one. We have also analysed the case of a potential (\ref{eq:pot:inflectionPoint:cubic}) with a flat inflection point, $V\propto 1+(\phi/\phi_0)^3$, where we have shown that ultra-slow roll is stable in the range $0<\phi<\Mp$, \ie when approaching the inflection point. When $\phi_0\ll\Mp$ the ultra-slow-roll regime is always short lived, but when $\phi_0\gg \Mp$ it can last for a large number of \efolds. 
In fact, if one considers for instance a potential of the form $V\propto 1+\alpha\ee^{\beta\phi/\Mp}$, with $\alpha\beta\ll 1$ and $\beta\gg 1$, there is even an infinite phase of classical ultra-slow roll for $\phi<0$.

In practice, in models with a long-lived USR epoch, the classical description may eventually break down when quantum fluctuations become more efficient at driving the field than its (decreasing) residual velocity. We have estimated when this happens, and shown that classical USR can still be long lived. It would however be interesting to study how the phase-space dynamics is modified in the presence of large stochastic diffusion in the USR regime and we plan to do so in the future.

When the quantum and classical evolutions become comparable then the resulting primordial density perturbations after inflation become large, which can lead to the formation of primordial black holes \cite{Hawking:1971ei,Carr:1975qj, GarciaBellido:1996qt}. In such cases we need a non-perturbative formalism to describe the cosmological evolution on large scales, such as stochastic inflation \cite{Starobinsky:1986fx,Vennin:2015hra,Pattison:2017mbe,Biagetti:2018pjj}. Inflation models similar to the inflection point potential discussed in this work have been investigated as models for the origin of primordial black holes~\cite{Ezquiaga:2018gbw, Ozsoy:2018flq} and we plan to study the evolution nature of perturbations in such models in future work.
\acknowledgments
We kindly thank H\'ector Ram\'irez and Bob Scherrer for helpful discussions and comments. HA, VV and DW acknowledge support from the UK Science and Technology Facilities Council grant ST/N000668/1. VV acknowledges funding from the European Union's Horizon 2020 research and innovation programme under the Marie Sk\l odowska-Curie grant agreement N${}^0$ 750491.
\bibliographystyle{JHEP}
\bibliography{USR}
\end{document}